\documentclass[a4paper,fleqn]{cas-dc}

\usepackage[numbers]{natbib}

\usepackage{caption} 
\usepackage{graphicx}
\usepackage{amssymb}
\usepackage{subfigure}
\usepackage{soul}

\newcommand{\figref}[1]{Fig.~\ref{#1}}
\newcommand{\tblref}[1]{Table.~\ref{#1}}
\newcommand{\eqtref}[1]{Eq.~\ref{#1}}
\newcommand{\dgli}[1]{\textcolor[rgb]{0,0,0}{#1}}
\newcommand{\lzh}[1]{\textcolor[rgb]{0,0,0}{#1}}
\newcommand{\cmmnt}[1]{}

\def\tsc#1{\csdef{#1}{\textsc{\lowercase{#1}}\xspace}}
\tsc{WGM}
\tsc{QE}

\shorttitle{Modeling and Analysis on Efficiency Degradation of Lithium-ion Batteries }    

\shortauthors{short author list for running head}  

\title [mode = title]{Modeling and Analysis on Efficiency Degradation of Lithium-ion Batteries}  

\tnotemark[1] 

\tnotetext[1]{tnote text} 

\author[must]{Zihui Lin}
\fnmark[1]

\credit{<Credit authorship details>}

\affiliation[must]{organization={Macau University of Science and Technology},
            addressline={Avenida Wai Long, Taipa, Macau, China}, 
            city={Macau},
            postcode={0000}, 
            state={Macao},
            country={China}}

\author[must]{Dagang Li} \corref{cor1}

\ead{dgli@must.edu.cn}

\credit{}

\cortext[cor1]{Corresponding author}

\begin{abstract}
Efficiency of Battery Energy Storage Systems (BESSs) is increasingly critical as renewable energy generation becomes more prevalent on the grid. Therefore, it is necessary to study the energy efficiency of lithium-ion batteries, which are typically used in BESSs. The purpose of this study is to propose the State of Efficiency (SOE) as a measure of how efficiently batteries transfer energy, and to analyze what factors affect the SOE of a battery throughout its lifetime. Using NASA's data set, we measure the SOE of Nickel-Cobalt-Aluminum (NCA) lithium-ion batteries by calculating the ratio of energy generated and consumed during discharge and charge phases. A linear trend was observed in the SOE trajectories, which is confirmed by the Mann-Kendall (MK) trend test. Following that, a linear SOE degradation model was presented. Further analysis shows that ambient temperature, discharge current, and cutoff voltage all affect SOE in different ways. Using the SOE and its behavior observed in this study, Battery Management Systems (BMS) can improve the energy efficiency of batteries by adjusting operating conditions or developing better management strategies.
\end{abstract}
\begin{document}

\begin{highlights}
\item State of Efficiency (SOE) is proposed as a performance indicator of lithium-ion battery in terms of energy efficiency.
\item The degradation trajectory of SOE for NCA lithium-ion batteries is studied and a linear model is proposed to describe SOE degradation trend.
\item A number of factors \lzh{that affect} SOE have been identified and studied, including ambient temperature, discharge current, and cutoff voltage.
\item \lzh{We found NCA lithium-ion batteries have very weak memory effect on SOE under changing operating conditions, which is helpful when designing an efficient BESS.}
\end{highlights}

\maketitle


\section{Introduction}
\label{sec:sec1}

Unlike traditional power plants, renewable energy from solar panels or wind turbines needs \lzh{storage solutions} 
, such as BESSs to become reliable energy sources and provide power on demand \cite{olabi2022renewable}.
The lithium-ion battery, which is used as a promising component of BESS \cite{choi2021li} that are intended to store and release energy, has a high energy density and a long energy cycle life \cite{li2022battery}. \dgli{The performance of} lithium-ion batteries has a direct impact on both the BESS and renewable energy sources since a reliable and efficient power system must always match power generation and load \cite{koohi2020review}.

However, battery's performance can be affected by a variety of operating conditions \cite{hausbrand2015fundamental}, and its performance continuously degrades during usage. The degradation can be classified into two parts: when the equipment is at rest (static) and when it is in use (operational) \cite{wohlfahrt2004aging}. In a static condition, the factors that determine degradation are temperature, State of Charge (SOC), and rest time, whereas during dynamic operation, the factors that determine degradation are rate (power), depth of discharge, and heat \cite{goodenough2010challenges, birkl2017degradation}. 

Regarding lithium-ion battery as an energy storage device, most studies are currently focused on its capacity management, charging rate, and cycle times \cite{tarascon2001issues}. A BMS of a BESS typically manages the lithium-ion batteries' State of Health (SOH) and Remaining Useful Life (RUL) in terms of capacity (measured in ampere hour) \cite{tarascon2001issues}. As part of the management of BESS, it has been possible to estimate and predict SOH and RUL. SOH is a key indicator for the estimation of battery life by focusing primarily on the maximum capacity that the battery is currently capable to supply. SOC indicates the capacity that the battery currently can provide at the present time. Estimating the SOC can provide insight into the battery's current capacity, while 
the SOH trajectory can help predict the battery's life regarding its capacity. 
Despite the fact that the battery's capacity is one of the most critical performance indicators, little attention has been paid to the battery's 
efficiency as it ages.

The efficiency of lithium-ion batteries greatly affects the efficiency of BESSs, which should minimize energy loss during operations.
This 
becomes increasingly important
when more renewable energy sources are connected to the grid and handled by BESSs \cite{choi2021li}. 
As the core energy storage component, if the batteries can only return an unexpected small fraction of the energy that is used to charge them, that will become a big problem for both the BESSs and the \dgli{power grid as a whole}.

Coulombic Efficiency (CE) \cite{hobold2021moving} was used as an indicator of lithium-ion battery efficiency in the reversibility of electrical current \cite{xiao2020understanding}, which actually has a direct relationship with the battery's capacity \cite{yang2018study}. It should be noted, however, that capacity and energy are not equivalent. Since the energy levels of lithium-ions are different during the redox reaction, regeneration requires more electromotive force than discharge due to the different voltage levels  \cite{eftekhari2017energy}. Therefore, even if lithium-ion battery has a high CE, it may not be energy efficient.

Battery's energy efficiency is critical to the sustainability of the planet due to its broad application \cite{pilkington2011relative}. Researchers have investigated BESS applications to maintain an acceptable level of efficiency \cite{qian2010high}. A study has also been conducted on the energy efficiency of electric vehicles \cite{wu2015electric}. The optimal battery SOC range for battery vehicles has been estimated by analyzing the relationship between battery energy efficiency and battery SOC range \cite{redondo2017impact}. There has also been discussion of the energy efficiency of NiMH batteries as they are applied to grid frequency control at various discharge currents \cite{zurfi2016experimental}. To the best of the author's knowledge, 
\dgli{there is still no study yet focusing on what factors will, and how they affect the energy efficiency of lithium-ion batteries in the long run across their entire life cycle.}

The following are the contributions of this study:
\begin{itemize}
\item State of Efficiency (SOE) is proposed as a performance indicator of lithium-ion battery in terms of energy efficiency.
\item The degradation trajectory of SOE for NCA lithium-ion batteries is studied and a linear model is proposed to describe SOE degradation trend.
\item A number of factors \lzh{that affect} SOE have been identified and studied, including ambient temperature, discharge current, and cutoff voltage.
\item \lzh{We found NCA lithium-ion batteries have very weak memory effect on SOE under changing operating conditions, which is helpful when designing an efficient BESS.}
\end{itemize}

The remainder of this paper is organized as follows. In Section 2, the SOE is presented along with its relationship to CE and SOH, as well as the adopted data set. In Section 3, the SOE trajectories of batteries is analyzed in a variety of application scenarios. We verify the linear relationship between SOE and cycle number by using time series analysis, and present the SOE degradation trend model and its regression results. Finally, in Section 4, we discuss the phenomena caused by changes in ambient temperature, discharge current, and cut-off voltage separately.



\section{Energy efficiency of lithium-ion battery}
\label{sec:sec2}
\subsection{The NASA \lzh{battery degradation} data set}

For this study, we use data set which is provided by the Prognostics Center of Excellence (PCoE) at NASA Ames \cite{bds2017nasa}. \lzh{In these experiments \cite{goebel2008prognostics}, 18650-size NCA lithium-ion batteries continue to work until End of Life (EoL) is reached, thereby providing data over battery lifespan. This is helpful to see how energy efficiency has changed over time. }

\lzh{These experiments, as shown in \tblref{tbl:1}, are designed to reduce battery performance in a controlled way. The test scheme,  as shown in \tblref{tbl:2}, is with a test matrix that included four cut-off voltages (2.7V, 2.5V, 2.2V and 2.5 V), three discharge currents (1A, 2A, and 4A), four ambient temperatures (4°C, 24°C, 43°C), and two stopping criteria (20\% and 30\% capacity fade). In most cases, an individual battery's operating conditions are different from those of other batteries, and remain constant throughout the test. As a result, this data set captures battery degradation data under various scenarios, which is suitable for analyzing battery efficiency under a wide range of operating conditions.}

\begin{table}[ht]
\caption{Technical indicators of experimental lithium-ion batteries.}\label{tbl:1}
\begin{tabular}{ll}
\specialrule{0.05em}{3pt}{3pt}
{Manufacturer}                 & {Idaho National Laboratory} \\
\hline
{Type}                         & {18650}                     \\
\hline
{Cathode/Anode}                & {NCA/Graphite}              \\
\hline
{Rated capacity/Rated voltage} & {2 Ah/3.7 V}                \\
\hline
{charging currents}            & {1A for CC stage}           \\
\hline
{Cutoff voltages}             & {4.2 V to 2/2.2/2.5/2.7 V}  \\
\hline
{Discharging currents}         & {1/2/4A}                    \\
\hline
{Ambient temperature}          & {4/24/43°C}                  \\
\hline
{stop criteria}                & {20/30\% capacity fade}    \\
\hline
\end{tabular}
\end{table}

\lzh{Each cycle, batteries are fully charged in CC-CV (Constant Current, Constant Voltage) mode and discharged in CC (Constant Current) mode to a certain depth, which is regulated by cut-off voltage. Throughout the cycle, voltage and current at the battery terminal are collected, which can be used to calculate how efficiently energy is reproduced in the specific cycle. }

\begin{table*}[htbp]
\centering
\caption{Test scheme of batteries.}\label{tbl:2}
\setlength{\tabcolsep}{5.7mm}{
\begin{tabular}{|c|cc|cc|c|c|}
\hline
\multicolumn{1}{|l|}{{Ambient Temperature}}                                                          & \multicolumn{2}{c|}{{4°C}}                                   & \multicolumn{2}{c|}{{24°C}}                                  & {43°C}   & {24°C and 44°C}   \\ \hline
\multicolumn{1}{|l|}{{\begin{tabular}[c]{@{}l@{}}Discharge Current \\ Cut-off Voltage\end{tabular}}} & \multicolumn{1}{c|}{{1A}}    & {2A}    & \multicolumn{1}{c|}{{2A}}    & {4A}    & {4A}    & {1A, 2A, and 4 A} \\ \hline
{2.0V}                                                                                               & \multicolumn{1}{c|}{{B0045}} & {B0053} & \multicolumn{1}{c|}{{}}      & {B0033} & {B0029} & {}              \\ \hline
{2.2V}                                                                                               & \multicolumn{1}{c|}{{B0046}} & {B0054} & \multicolumn{1}{c|}{{B0007}} & {B0034} & {B0030} & {B0038}         \\ \hline
{2.5V}                                                                                               & \multicolumn{1}{c|}{{B0047}} & {B0055} & \multicolumn{1}{c|}{{B0006}} & {}      & {B0031} & {B0039}         \\ \hline
{2.7V}                                                                                               & \multicolumn{1}{c|}{{B0048}} & {B0056} & \multicolumn{1}{c|}{{B0005}} & {}      & {B0032} & {B0040}         \\ \hline
\end{tabular}}
\end{table*}

Additionally, the data set includes a group of batteries utilising multiple load current levels (1A, 2A, and 4A) as ambient temperature steps from 24°C to 44 °C. The study will use this group of batteries to investigate the characteristics of energy efficiency. Due to the particular characteristics of this group, the series of cycles has been divided into three continuous parts so that uniform data can be collected. 

According to some studies, a fresh battery's CE may exceed 100\% because the internal electrochemical characteristics are not stable in the initial cycles \cite{xiao2020understanding}. Furthermore, the documentation of the data set indicates that there are abnormalities due to software errors or other unknown causes, such as stopping the test before reaching the specified cut-off voltage. Considering that the purpose of this study is to determine the characteristics of lithium-ion battery's energy efficiency throughout its lifespan, as well as to identify the trend of SOE and the influence of relevant operating conditions on it, the first cycle as well as some special abnormal cycles for each battery have been excluded.

\subsection{State of efficiency}

\begin{figure}[ht]    
    \centering
    \includegraphics[scale=0.3]{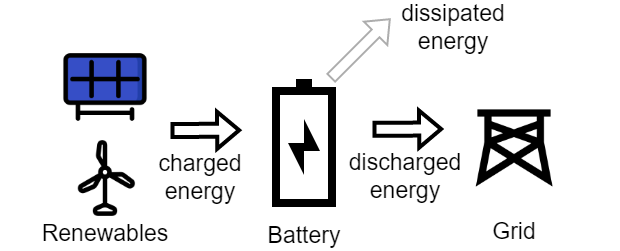}
    \caption{Energy conservation during battery charging and discharging.}
    \label{fig:1}
\end{figure}

As an energy intermediary, lithium-ion batteries are used to store and release electric energy. An example of this would be a battery that is used as an energy storage device for renewable energy. The battery receives electricity generated by solar or wind power production equipment. Whenever there is a demand from the grid, the stored electric energy is released. Inevitably, this process involves the dissipation of energy. As a result of polarization, the battery's energy dissipates during the charge-discharge process because coulomb losses from non-productive chemical side reactions and the battery's terminal voltage drops when current flows through  it \cite{zurfi2017electrical}. Therefore, while batteries are in operation, they lose energy during both cycle aging and calendar aging, and the amount varies depending on how they are used. \figref{fig:1} shows energy conservation in the application of battery. Energy dissipates during charging and discharging, and the energy in and out of the battery remains balanced.
\begin{equation} \label{eq:1}
    E_{charge}=E_{discharge}+E_{dissipate},
\end{equation}
where $ E_{discharge} $ and $ E_{dissipate} $ represent the energy released during discharge and the energy dissipated during battery operation. The dissipated energy in operation can be expressed as

\begin{equation} \label{eq:2}
    E_{dissipate}=E_{disCycle}+E_{disCalendar},
\end{equation}
where $ E_{disCycle} $ and $ E_{disCalendar} $ represent the dissipated energy during the cycle aging and calendar aging. \lzh{Existing researches have studied SOH, which refers to the maximum remaining capacity of a battery over its rated capacity, that is, the capacity performance of a battery.} In contrast to SOH, SOE focuses on the battery's efficiency in using energy, as \lzh{discharge energy in a battery is always less than charge energy.}

The USA PNGV battery test manual \cite{shhpngv} gives a \lzh{intuitive} definition of round-trip efficiency, but does not have a strict specific test protocol. PNGV round-trip efficiency is defined as
\begin{equation} \label{eq:3}
    \text{Round-trip Efficiency} = \frac{watt \cdot hours(discharge)}{watt \cdot hours(regen)} \times 100 \%.
\end{equation}

\lzh{Considering batteries are energy storage devices, we propose a calculation of battery efficiency based on the ratio of output energy to input energy.} As this study aims to evaluate the energy efficiency of a complete charging and discharging process, SOE is defined as

\begin{equation} \label{eq:4}
    SOE=\frac{E_{discharged}}{E_{charged}} , 
\end{equation}

where battery efficiency is calculated as the ratio between the amount of energy the battery can supply during discharge and the amount of energy it consumes during charging. The charged energy is the accumulation of power during charging:

\begin{equation} \label{eq:5}
    E_{charged}=\int_{t_0}^{t_n}P_{charge}(t)dt=\int_{t_0}^{t_n}V_{charge}(t)I_{charge}(t)dt ,
\end{equation}
where $ P_{charge} (t) $ is the charging power as a function of time, which is obtained by multiplying $ V_{charge} (t) $ and $ I_{charge} (t) $, that is, the product of voltage and current at the terminal of the battery. It is common practice in industrial applications to sample discretely, then \eqtref{eq:5} can be simplified as

\begin{equation} \label{eq:6}
    E_{charged}=\sum_{i=0}^{n-1}{P_{charge}\Delta t}=\sum_{t=0}^{n-1}{V_{charge}I_{charge}\Delta t_i} ,
\end{equation}

Similarly, we can also measure the energy during discharge. Hence, in this study, SOE is evaluated as

\begin{equation} \label{eq:7}
    SOE=\frac{E_{discharged}}{E_{charged}}=\frac{\sum_{i=0}^{n-1}{V_{discharge}I_{discharge}\Delta t_i}}{\sum_{i=0}^{n-1}{V_{charge}I_{charge}\Delta t_i}}.
\end{equation}

where $ V_{discharge} $, $ I_{discharge} $, $ V_{charge} $ and $ I_{charge} $ are provided in the data set and $ \Delta t_i $ is calculated by their timestamp.

\subsection{Coulombic efficiency and energy efficiency}

CE as a battery parameter to monitor the magnitude of side reactions \cite{yang2018study} is defined as

\begin{equation} \label{eq:8}
    CE = \frac {C_{discharge}}{C_{charge}},
\end{equation}
where $ C_{discharge} $ is the discharge capacity of a battery at a single cycle, and $ C_{charge} $ is the charge capacity of the battery in the same cycle. By definition, CE is the ratio of discharge capacity over charge capacity of a lithium-ion battery. Because capacity is measured by the total charge flow to/from the electrode and the total capacity of lithium-ion batteries is usually cathode-limiting, CE can be expressed as the ratio between the amount of lithium-ions or electrons returning to cathode and the amount of lithium-ions or electrons departing from cathode in a full cycle, as indicated in \eqtref{eq:8}.

A battery's CE refers to its ability to regenerate Li+ or electrons. In an ideal battery, where there are no side reactions at the electrodes, lithium-ions or electrons should flow solely as a result of reversible electrochemical reactions with CE equal to 100\%. Realistic batteries, however, exhibit numerous side reactions between the electrode and the electrolyte. Due to the presence of irreversible side reactions in the battery, the CE is always less than 100\%. Generally, modern lithium-ion batteries have a CE of at least 99.99\% if more than 90\% capacity retention is desired after 1000 cycles \cite{xiao2020understanding}.

\begin{figure}[htbp]
	\centering  
	\subfigcapskip=-5pt 
	\subfigure[100\% energy efficiency]{
		\includegraphics[width=0.48\linewidth]{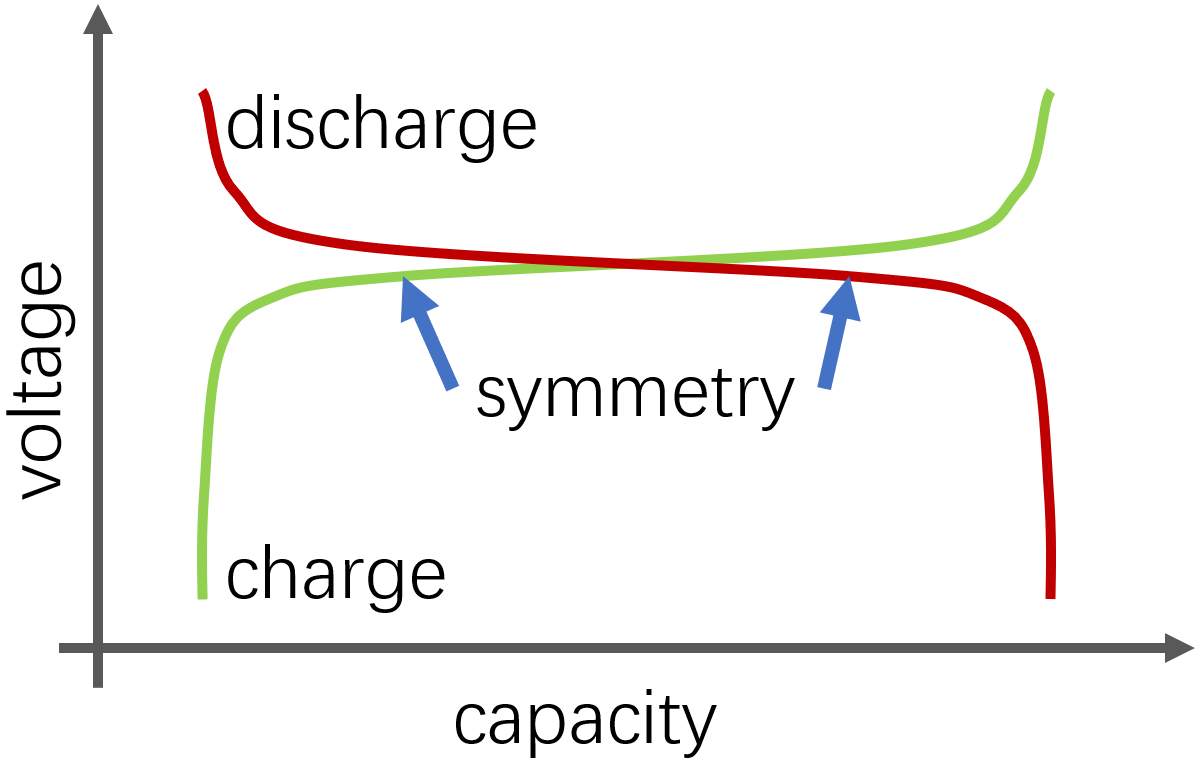}}
	\subfigure[90\% energy efficiency due to overpotential]{
		\includegraphics[width=0.48\linewidth]{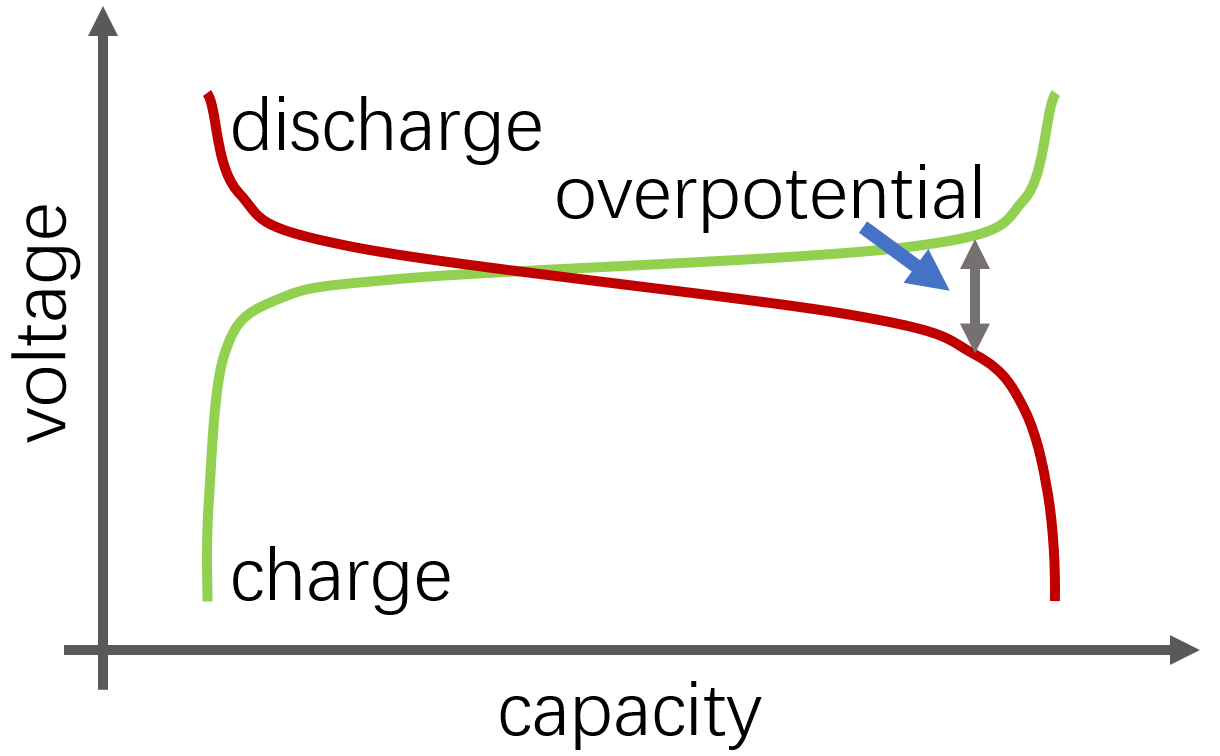}}
	\subfigure[50\% lower energy efficiency, resulting in energy waste]{
		\includegraphics[width=0.48\linewidth]{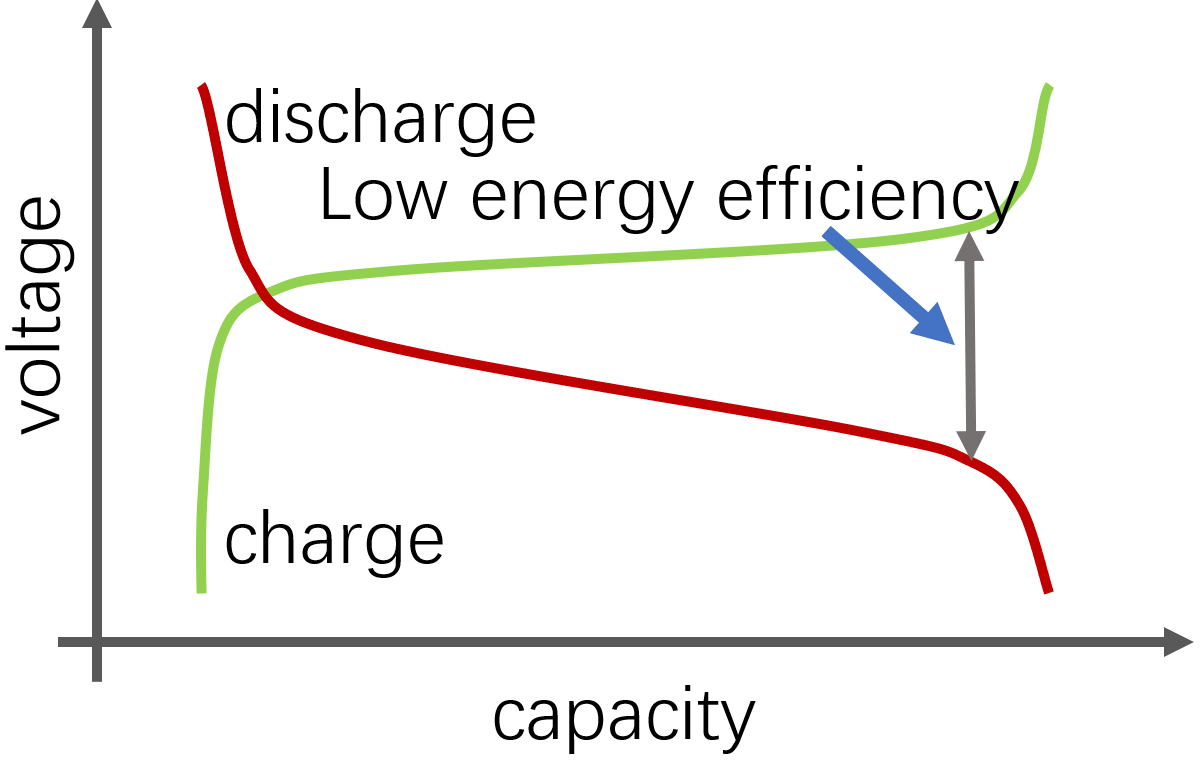}}
	
	\caption{Three different scenarios \cite{eftekhari2017energy} at 100\% CE.}
        \label{fig:2}
\end{figure}

However, the coulombic efficiency of a battery cannot be equated with its energy efficiency. This is due to the fact that lithium atoms participate in redox reactions at different energy levels. The overpotential that is consumed during charging cannot be recovered during discharging. Due to the internal resistance of the battery, some energy must be consumed to overcome the resistance, which limits the battery's ability to achieve 100\% energy efficiency \cite{eftekhari2017energy}.

\figref{fig:2} illustrates the charging/discharging process with both 100\% CE, corresponding to different electrical energy efficiencies. For the ideal 100\% energy efficiency in (a), the charge/discharge curves are perfectly symmetrical, meaning that the stored lithium-ions have the same energy level as in both the charge and discharge phases. Nevertheless, the electrons of the charge phase have a higher energy level than the electrons of the discharge phase in (b), which will result in an energy efficiency of only 90\%. (c) represents a very low energy conversion efficiency of the battery, where 50\% of the energy is wasted in the charging/discharging process despite the fact that 100\% of capacity is available.

\subsection{SOH versus SOE}

Numerous studies have been conducted on SOH for energy storage batteries. Served as a measure of the battery's aging process, SOH is commonly calculated as

\begin{equation} \label{eq:9}
    SOH=\frac{Q_{max}}{Q_{rated}},
\end{equation}
where $ Q_{max} $ and $ Q_{rated} $ represent the battery's maximum discharge capacity with its rated capacity. RUL is defined by the number of cycles a battery can undergo before its capacity deteriorates and reaches EoL which is typically defined as when the battery has only 80\% of its rated capacity left. 

\begin{figure}[htbp]
	\centering  
	\subfigcapskip=-5pt 
	\subfigure[B0005 discharged to 2.7V at 24 °C and 2A]{
		\includegraphics[width=0.9\linewidth]{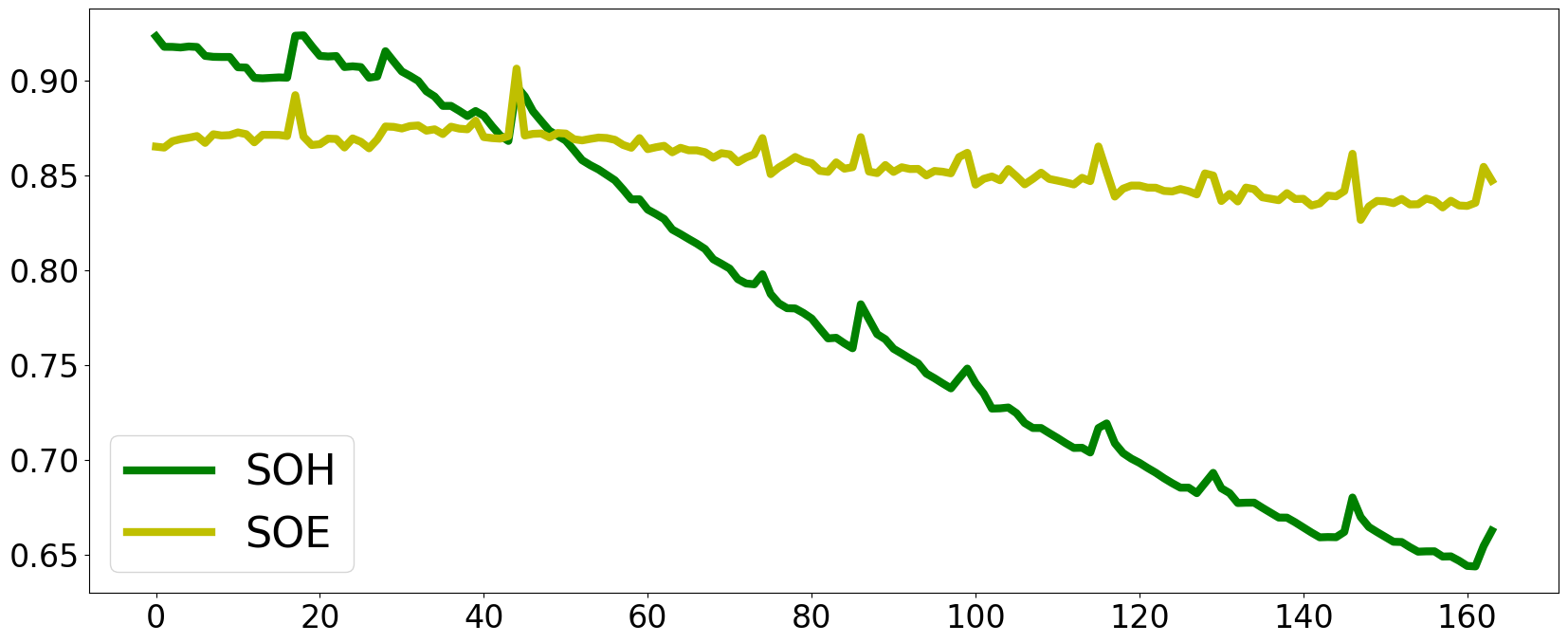}}
	\subfigure[B00029 discharged to 2.0V at 43 °C and 4A]{
		\includegraphics[width=0.9\linewidth]{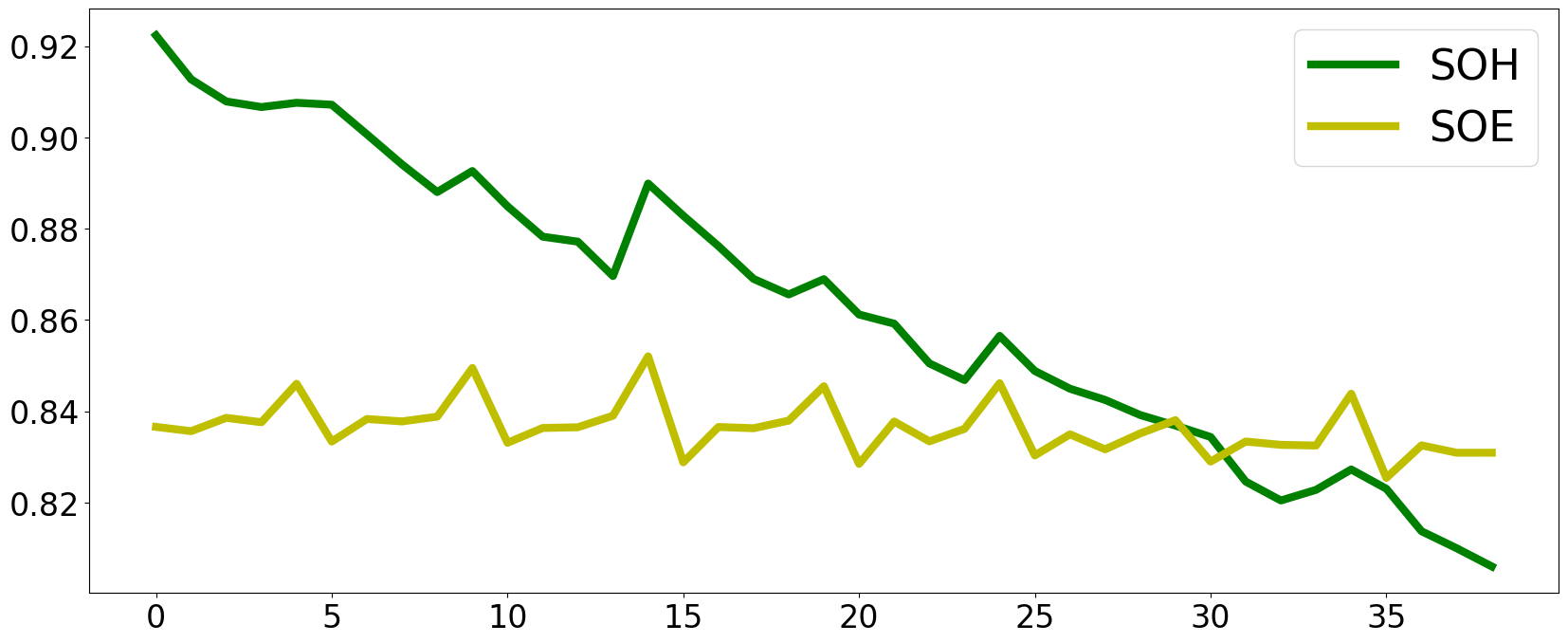}}
        \subfigure[B00033 discharged to 2.0V at 24 °C and 4A]{
		\includegraphics[width=0.9\linewidth]{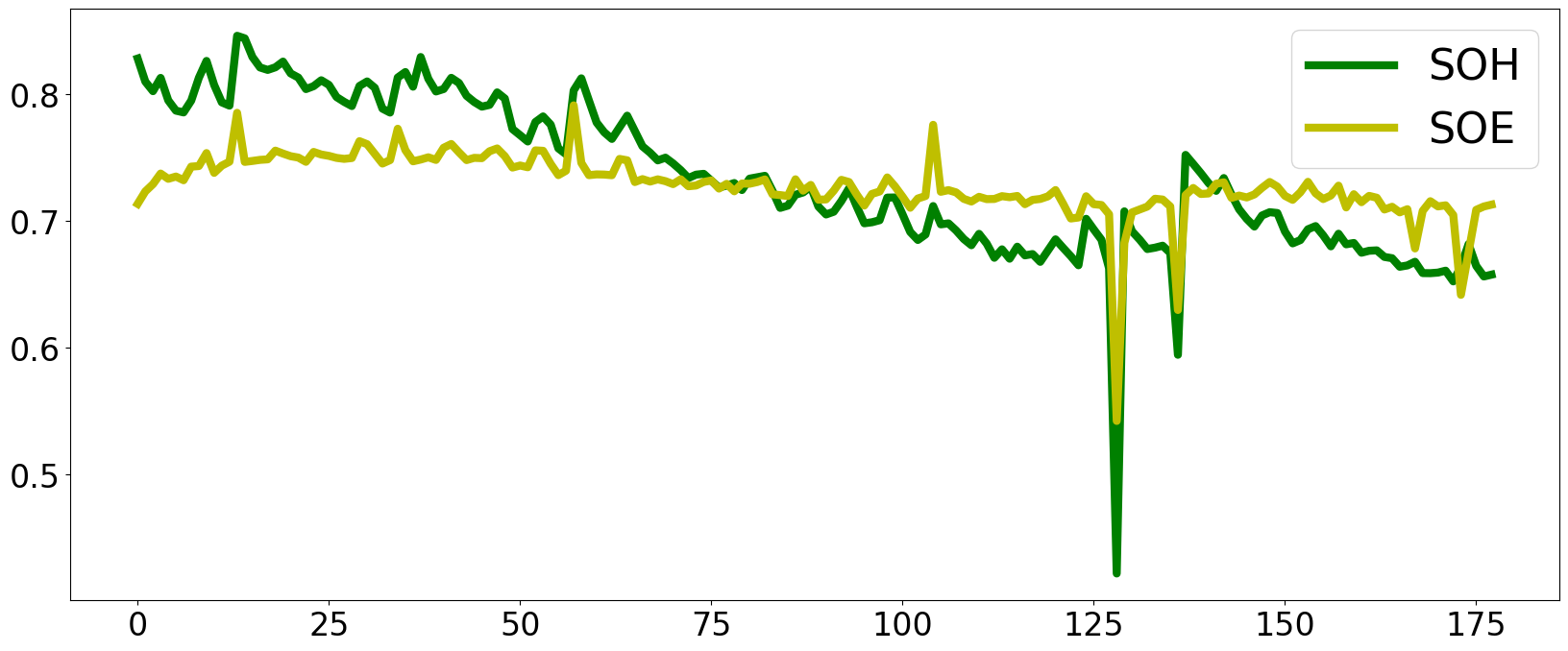}}
	\subfigure[B0045 discharged to 2.0V at 4 °C and 1A]{
		\includegraphics[width=0.9\linewidth]{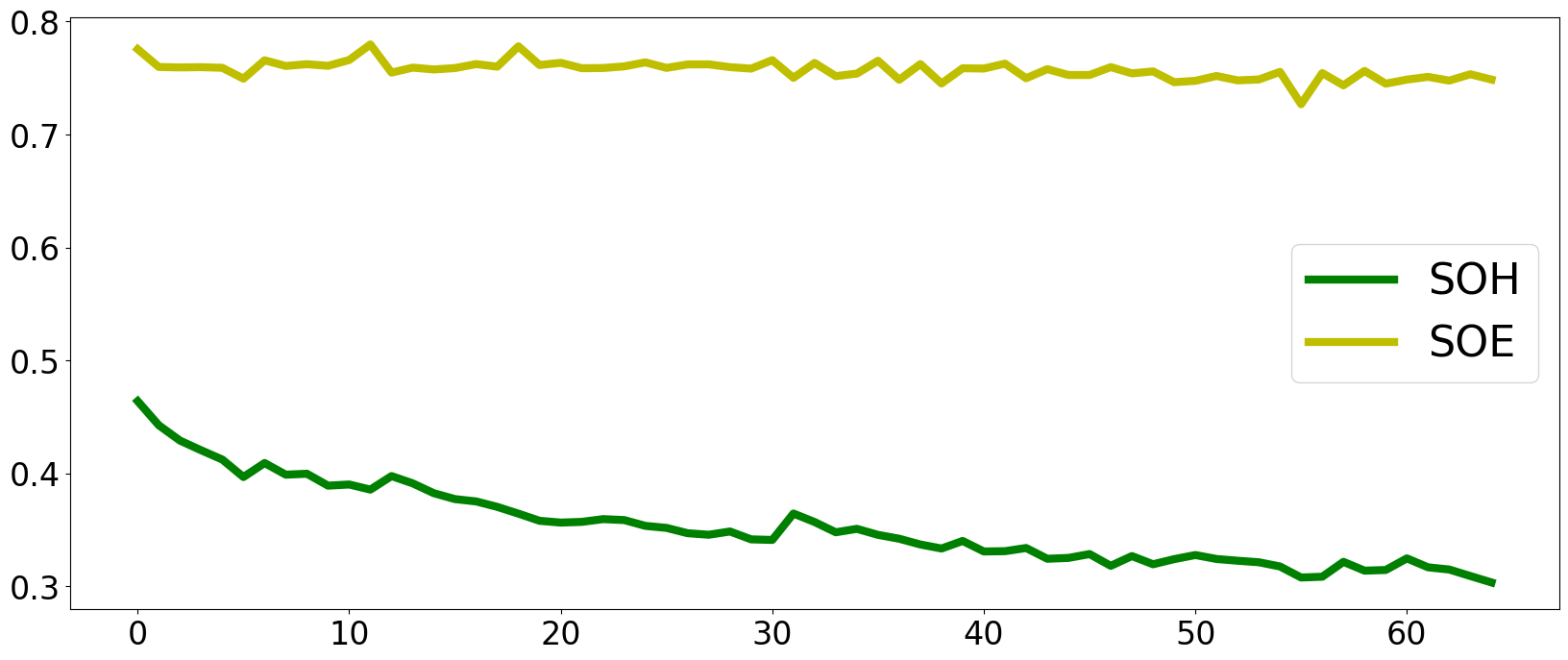}}
	
	\caption{Typical SOH and SOE trajectories over cycling.}
        \label{fig:3}
\end{figure}

\figref{fig:3} illustrates typical SOE and SOH trends as the battery ages over a number of cycles. Due to the wide range of use scenarios included in the data set, including different discharge intensities and depths in various ambient temperatures, these batteries exhibit distinctly different trajectories in terms of their SOH and SOE. As an example, when the battery B0005 is tested at an ambient temperature of 24°C, with a discharge current of 2A and a cutoff voltage of 2.7V, its SOE is essentially above 0.83 while its SOH is continuously reduced to less than 0.70. As a comparison, B0033 was also evaluated at 24°C, but is subjected to a higher intensity (4A current) and a deeper (2.0V cutoff voltage) discharge, it's SOE is around 0.7. The efficiency of B0005 was 0.86 when its capacity had decayed to about 0.8, whereas B0033's SOE was only 0.73 with similar SOH. In spite of the high discharge temperature of 43°C, the SOH of B0029 drops very rapidly, but it is almost always above 0.83 in terms of SOE. One of the extreme cases was the testing of B0045 at 4°C. Even though its SOH was very low (less than 0.5 throughout the test), it had a relatively high efficiency of about 0.78. As illustrated in these examples, SOH and SOE differ in how they explain battery performance. Generally, SOH describes the health of a battery in terms of its ability to release Coulombs. While SOE describes the efficiency of a battery as an energy storage medium in terms of the ratio of energy transfer during charging and discharging.

\begin{table}[htbp]
\caption{Typical SOE and SOH values.}\label{tbl:3}
\setlength{\tabcolsep}{8mm}{
\begin{tabular}{lll}
\specialrule{0.05em}{3pt}{3pt}
Batteries & SOH              & SOE              \\
B0005        & 90.23\%          & 87.59\%          \\
B0039                                                  & 88.58\%          & 90.50\% \\
B0005                                                  & 80.08\%          & 86.10\% \\
B0033                                                  & 81.28\%          & 73.73\% \\
B0005                                                  & 70.06\%          & 84.45\% \\
B0045                                                  & 46.40\%          & 77.55\%
\end{tabular}}
\end{table}

To quantitatively analyze the correlation between SOE and SOH, we used the person correlation coefficient (PCC). the PCC is calculated as follows:
\begin{equation} \label{eq:10}
    PCC = \frac{\sum(SOE - \overline{SOE})(SOH - \overline{SOH})}
            {\sqrt{\sum(SOE-\overline{SOE})^{2}\cdot\sum(SOH-\overline{SOH})^{2}}}\\.
\end{equation}

\begin{table}[ht]
\caption{PCC of SOH and SOE.}\label{tbl:4}
\setlength{\tabcolsep}{2.6mm}{
\begin{tabular}{lllll}
\specialrule{0.05em}{3pt}{3pt}
\cellcolor[HTML]{DED1CA}{\begin{tabular}[c]{@{}l@{}}0.5406\\ (B0045)\end{tabular}} & \cellcolor[HTML]{F8CCAE}{\begin{tabular}[c]{@{}l@{}}0.9218\\ (B0046)\end{tabular}} & \cellcolor[HTML]{ECCEBA}{\begin{tabular}[c]{@{}l@{}}0.7705\\ (B0047)\end{tabular}} & \cellcolor[HTML]{E5CFC2}{\begin{tabular}[c]{@{}l@{}}0.6512\\ (B0048)\end{tabular}} & \cellcolor[HTML]{BDD7EE}{\begin{tabular}[c]{@{}l@{}}0.0333\\ (B0053)\end{tabular}} \\
\cellcolor[HTML]{BED7ED}{\begin{tabular}[c]{@{}l@{}}0.0502\\ (B0054)\end{tabular}} & \cellcolor[HTML]{BDD7EE}{\begin{tabular}[c]{@{}l@{}}0.0254\\ (B0055)\end{tabular}} & \cellcolor[HTML]{C3D6E7}{\begin{tabular}[c]{@{}l@{}}0.1265\\ (B0056)\end{tabular}} & \cellcolor[HTML]{F8CBAD}{\begin{tabular}[c]{@{}l@{}}0.9442\\ (B0007)\end{tabular}} & \cellcolor[HTML]{F6CCAF}{\begin{tabular}[c]{@{}l@{}}0.9161\\ (B0006)\end{tabular}} \\
\cellcolor[HTML]{F0CDB5}{\begin{tabular}[c]{@{}l@{}}0.8326\\ (B0005)\end{tabular}} & \cellcolor[HTML]{F6CCAF}{\begin{tabular}[c]{@{}l@{}}0.9183\\ (B0033)\end{tabular}} & \cellcolor[HTML]{E3D0C4}{\begin{tabular}[c]{@{}l@{}}0.6205\\ (B0034)\end{tabular}} & \cellcolor[HTML]{D6D2D2}{\begin{tabular}[c]{@{}l@{}}0.4282\\ (B0029)\end{tabular}} & \cellcolor[HTML]{C7D5E2}{\begin{tabular}[c]{@{}l@{}}0.1962\\ (B0030)\end{tabular}} \\
\cellcolor[HTML]{C2D6E9}{\begin{tabular}[c]{@{}l@{}}0.1075\\ (B0031)\end{tabular}} & \cellcolor[HTML]{D5D3D4}{\begin{tabular}[c]{@{}l@{}}0.3998\\ (B0032)\end{tabular}} & {}    & {}    & {}                                                                                
\end{tabular}}
\end{table}

As shown in the \tblref{tbl:4}, there is no obvious correlation between SOH and SOE during battery aging. However, it is a necessity to evaluate the efficiency of batteries in energy storage scenarios.

\section{Modeling State of Efficiency}
\subsection{Trend of SOE Trajectory}

\begin{figure}[htbp]    
    \centering
    \includegraphics[scale=0.25]{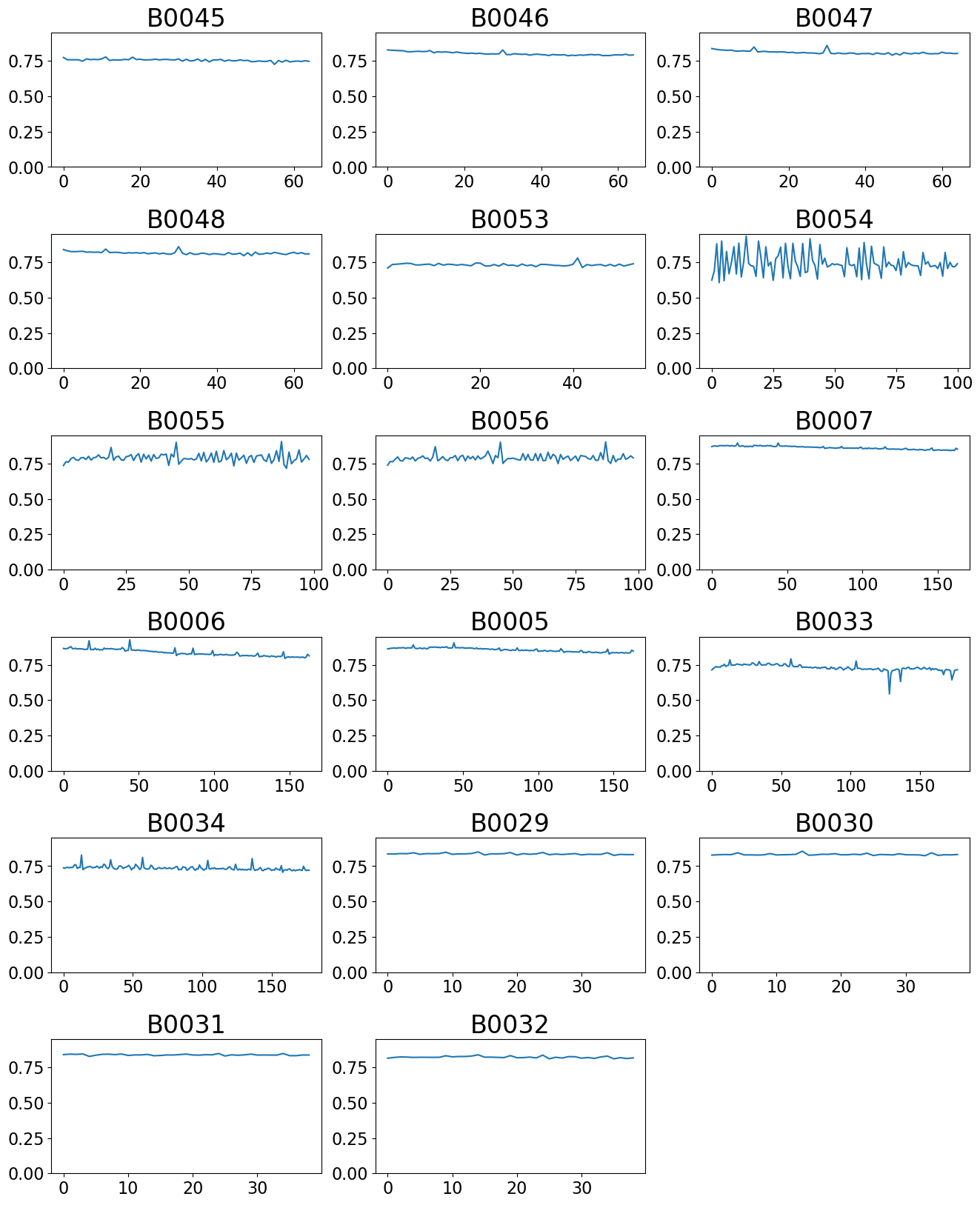}
    \caption{SOE trajectory under various operating conditions.}
    \label{fig:4}
\end{figure}

A battery undergoes a series of charging and discharging cycles during its aging process. For the data set used, the SOH of each battery decreases over time until it reaches the end-of-experiment condition (20\% or 30\% capacity fade). According to \eqtref{eq:7}, we calculate the SOE for each battery 
in each of its charging/discharging cycle. \figref{fig:4} shows the trajectory of 
SOE (between 0 and 1) over cycles for each battery during the aging process. These batteries' SOE trajectories all show a flat or slightly downward (decreasing) trend. In other words, 
\dgli{a somewhat linear trend is observed, despite different level of fluctuations is also present at different conditions.}

\subsection{Linearity of the SOE trajectory}

\dgli{In this subsection we want to test and prove that the SOE trajectory is indeed linear as we observed in the previous subsection. For this purpose, we first calculate the first difference of the SOE time series, which removes the first-order trend from the original trajectory; if the resulting first difference has not trend any more, then we can say that the original trajectory has a linear trend.}
The first difference for $ SOE_t $ is expressed as

\begin{equation} \label{eq:12}
\begin{aligned}
\Delta _{SOE_t}  & = SOE_t - SOE_{t-1} \\
\end{aligned},
\end{equation}

\lzh{To determine whether the $\Delta _{SOE}$ shows no trend,} we use The Mann-Kendall (MK) Trend Test.The MK Trend Test \cite{mann1945nonparametric, kendall1948rank} is used to analyze time series data for consistently increasing or decreasing trends.
MK trend test is also used in trend detection in hydrologic data\cite{hamed2008trend}.
For the time series $\Delta _{SOE}$ with the actual number of cycles performed in the experiment of n (sample size), the MK trend test first calculates S:

\begin{equation} \label{eq:13}
S=\sum_{k=1}^{n-1} \sum_{j=k+1}^{n} \operatorname{sgn}\left(\Delta _{SOE_{j}} - \Delta _{SOE_{k}}\right),
\end{equation}
where $ \operatorname{sgn}(x)=\left\{\begin{array}{rll}

1 & \text { if } & x>0 \\

0 & \text { if } & x=0 \\

-1 & \text { if } & x<0

\end{array}\right. $.

An MK test analyzes the sign of the difference between all data and its predecessors. Throughout the time series, each value is compared with its preceding value, giving $ n(n - 1) / 2 $ pairs of sign values. Intuitively, the number of sign values indicates the tendency of a trend's present. The variance of S is then calculated as

\begin{equation} \label{eq:14}
    \operatorname{VAR}(S)=\frac{1}{18}\left(n(n-1)(2 n+5)-\sum_{k=1}^{p} q_{k}\left(q_{k}-1\right)\left(2 q_{k}+5\right)\right),
\end{equation}
where p represents the number of tie groups of the time series SOE, and $ q_k $ represents the number of data of $ k $ tie groups. For the test statistic, transform S into $ Z_{MK} $ as follows:

\begin{equation} \label{eq:15}
Z_{M K}=\left\{\begin{array}{rll}
\frac{S-1}{V A R(S)} & \text { if } & S>0 \\
0 & \text { if } & S=0 \\
\frac{S+1}{V A R(S)} & \text { if } & S<0
\end{array}\right .
\end{equation}

Using the python package \cite{hussain2019pymannkendall} of MK trend test, the $ \Delta _ {SOE} $ series of each battery were examined and the results are shown in \tblref{tbl:5}. The z-table can be used to calculate the probability (p value). 

\begin{table}[htbp] 
\caption{The p values of Mann-Kendall trend test result.}\label{tbl:5}
\begin{tabular}{|l|l|l|l|l|}
\hline
{B0045}    & {B0046}    & {B0047}    & {B0048}    & {B0053}    \\ \hline
{0.767631} & {0.434133} & {0.689333} & {0.655519} & {0.957178} \\ \hline
{B0054}    & {B0055}    & {B0056}    & {B0005}    & {B0006}    \\ \hline
{0.700847} & {0.882874} & {0.662952} & {0.870048} & {0.579624} \\ \hline
{B0007}    & {B0033}    & {B0034}    & {B0029}    & {B0030}    \\ \hline
{0.774103} & {0.998988} & {0.698799} & {0.919888} & {0.687463} \\ \hline
{B0031}    & {B0032}    & {   -}     & {   -}     & {   -}        \\ \hline
{0.782101} & {0.880084} & {   -}     & {   -}     & {   -}        \\ \hline
\end{tabular}

\end{table}

Assuming a 5\% significance level, if the p value is less then 0.05, then the alternate hypothesis is accepted, which indicates the presence of a trend, whereas if the p value is greater than 0.10, then the null hypothesis is accepted, indicating the absence of a trend. \lzh{As the p value of results for these batteries in \tblref{tbl:5} are greater than 0.10, we reject the null hypothesis, and confirm that there's no consistently increasing or decreasing trend in the first difference of the SOE series. Therefore, the SOE trend of these batteries should obey a linear trend.}

\subsection{SOE linear model and regression results}

\lzh{As discussed earlier, the SOE trajectories of the NCA lithium-ion batteries all show a linear trend throughout their lifetime, we propose a linear SOE model, with the following formula:}

\begin{equation} \label{eq:11}
    SOE_t = \alpha t + \eta +\upsilon _t,
\end{equation}
where $ a $ is the slope of the trend, $ t $ is the cycle number, $ \eta $ is the energy efficiency of the battery at the beginning of the test, and $ \upsilon _t $ is the fluctuation around the linear trend. 

\lzh{As the fluctuations, which arise from stochasticity, are not dependent on the trend, we can represent the trend of SOE of a specific NCA lithium-ion battery under a certain operating condition} with two parameters: the slope $ \alpha $ and the initial value (the intercept) $ \eta $.

\begin{equation} \label{eq:16}
SOE_t = \alpha t + \eta .
\end{equation}

\begin{figure}[htbp]
	\centering  
	\subfigcapskip=-5pt 
	\subfigure[B0005 $\alpha$: -0.0003, $\eta$: 0.8784]{
		\includegraphics[width=0.48\linewidth]{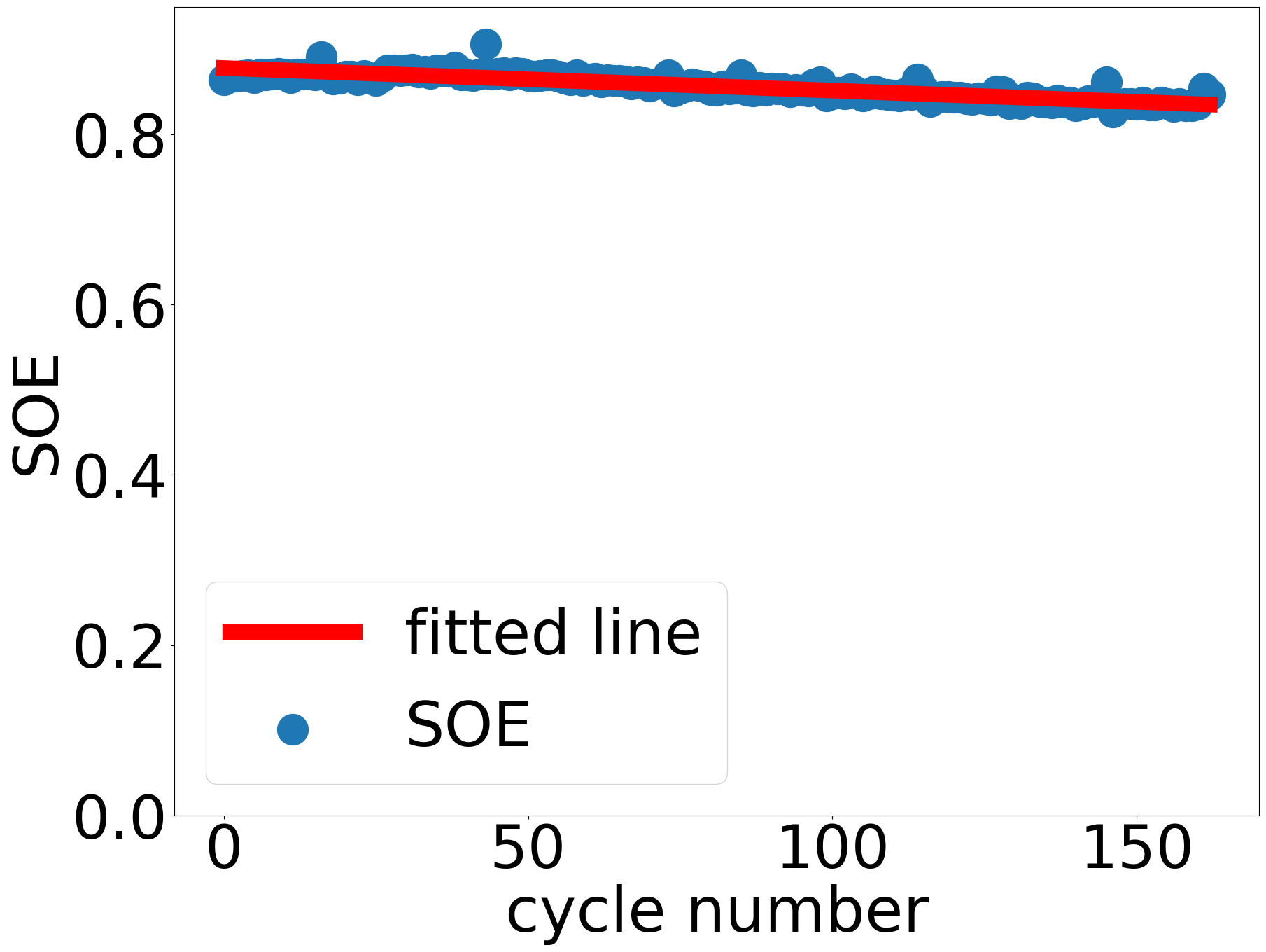}}
	\subfigure[B00029 $\alpha$: -0.0002, $\eta$: 0.8403]{
		\includegraphics[width=0.48\linewidth]{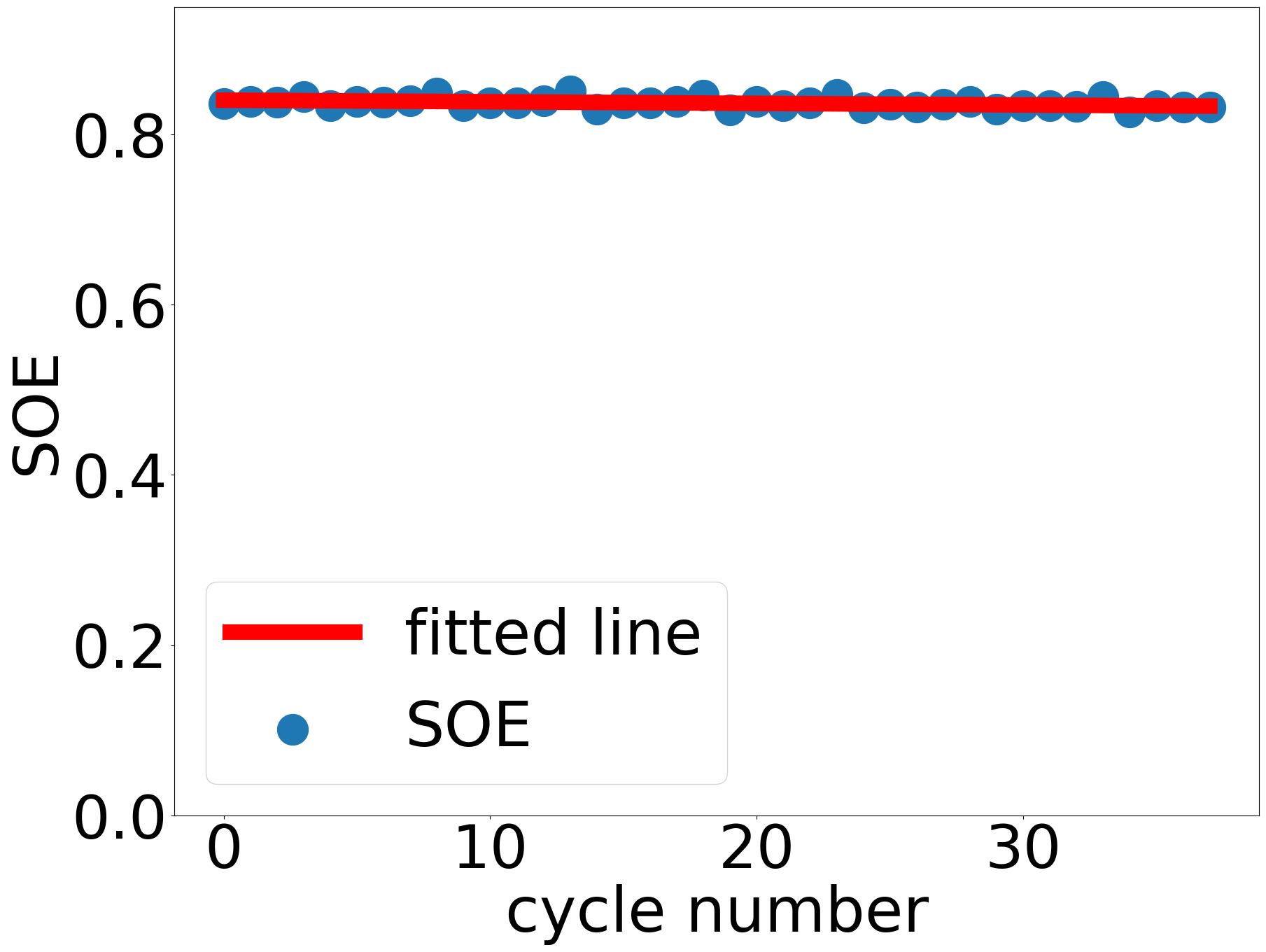}}
        \subfigure[B00033 $\alpha$: -0.0003, $\eta$: 0.7571]{
		\includegraphics[width=0.48\linewidth]{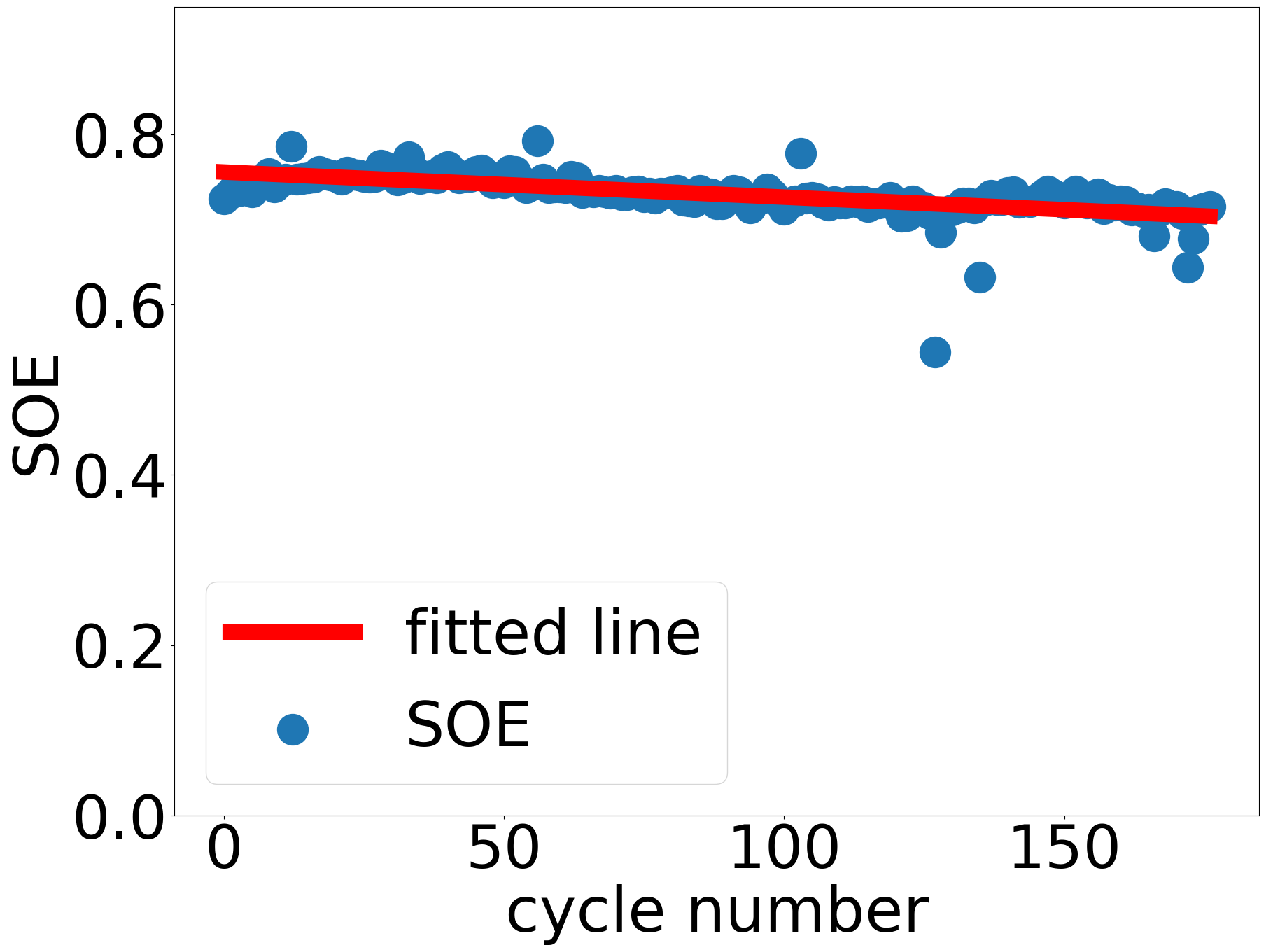}}
	\subfigure[B0045 $\alpha$: -0.0003, $\eta$: 0.7647]{
		\includegraphics[width=0.48\linewidth]{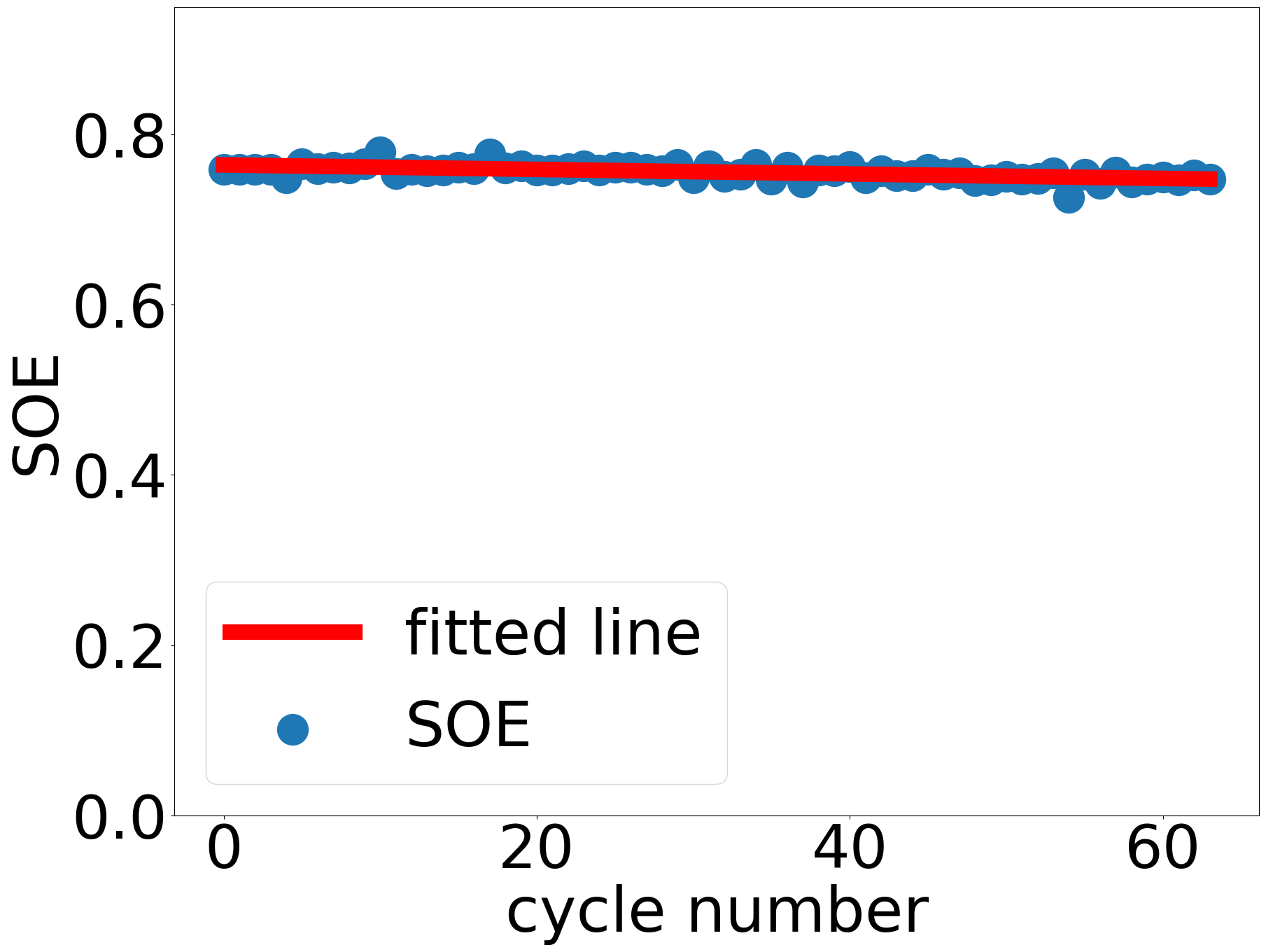}}
	
	\caption{Typical SOH and SOE trajectories over cycling.}
        \label{fig:5}
\end{figure}


Batteries in the data set that have been aged under various operating conditions. We use least squares regression to fit the SOE trajectories of every battery using SOE model presented as \eqtref{eq:16}. A selection of regression results are shown in \figref{fig:5}. 

\begin{figure}[htbp]
	\centering  
	\subfigcapskip=-5pt 
	\subfigure[SOE trends]{
		\includegraphics[width=0.95\linewidth]{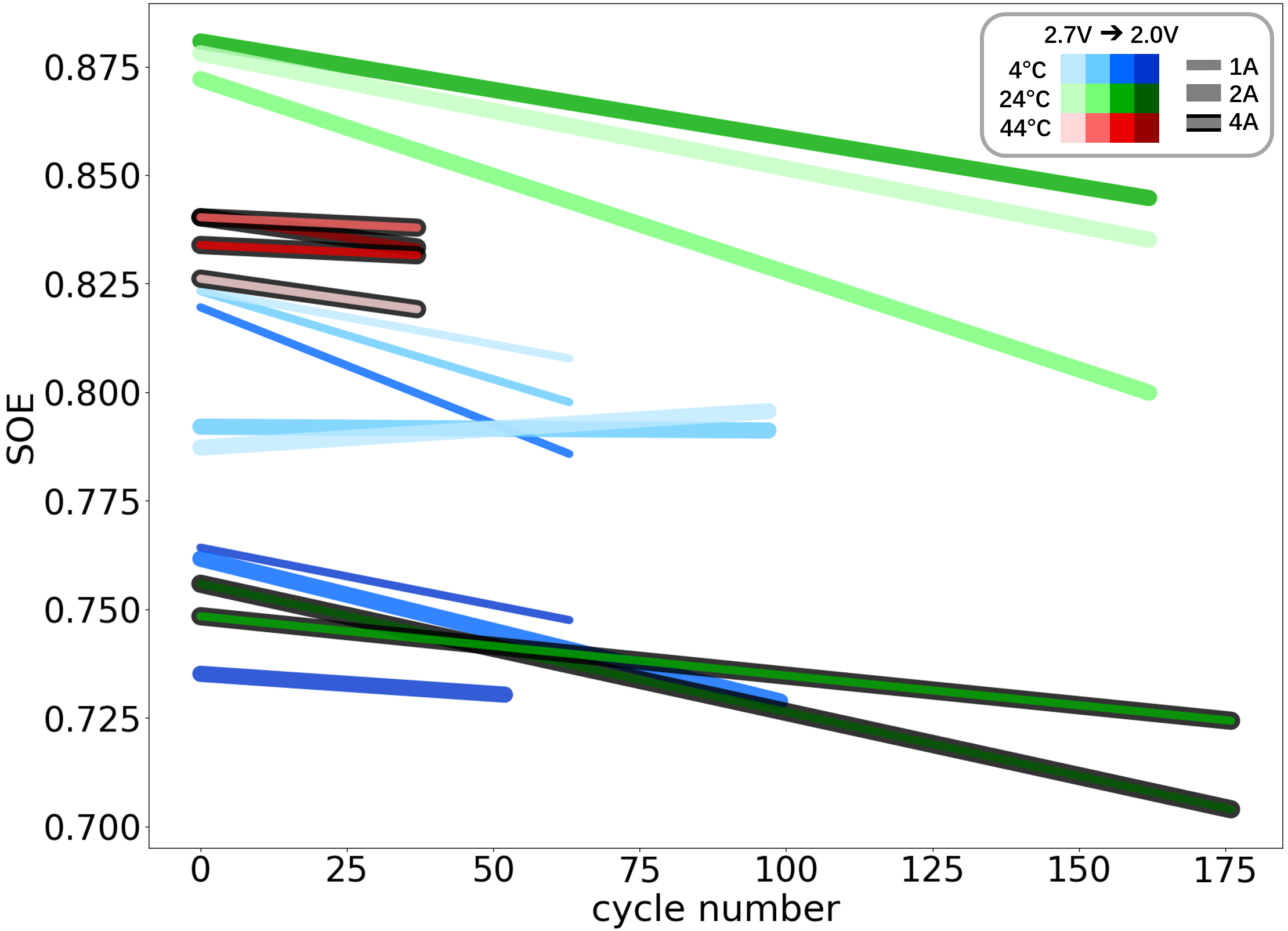}}
	\subfigure[SOE ranges]{
		\includegraphics[width=0.95\linewidth]{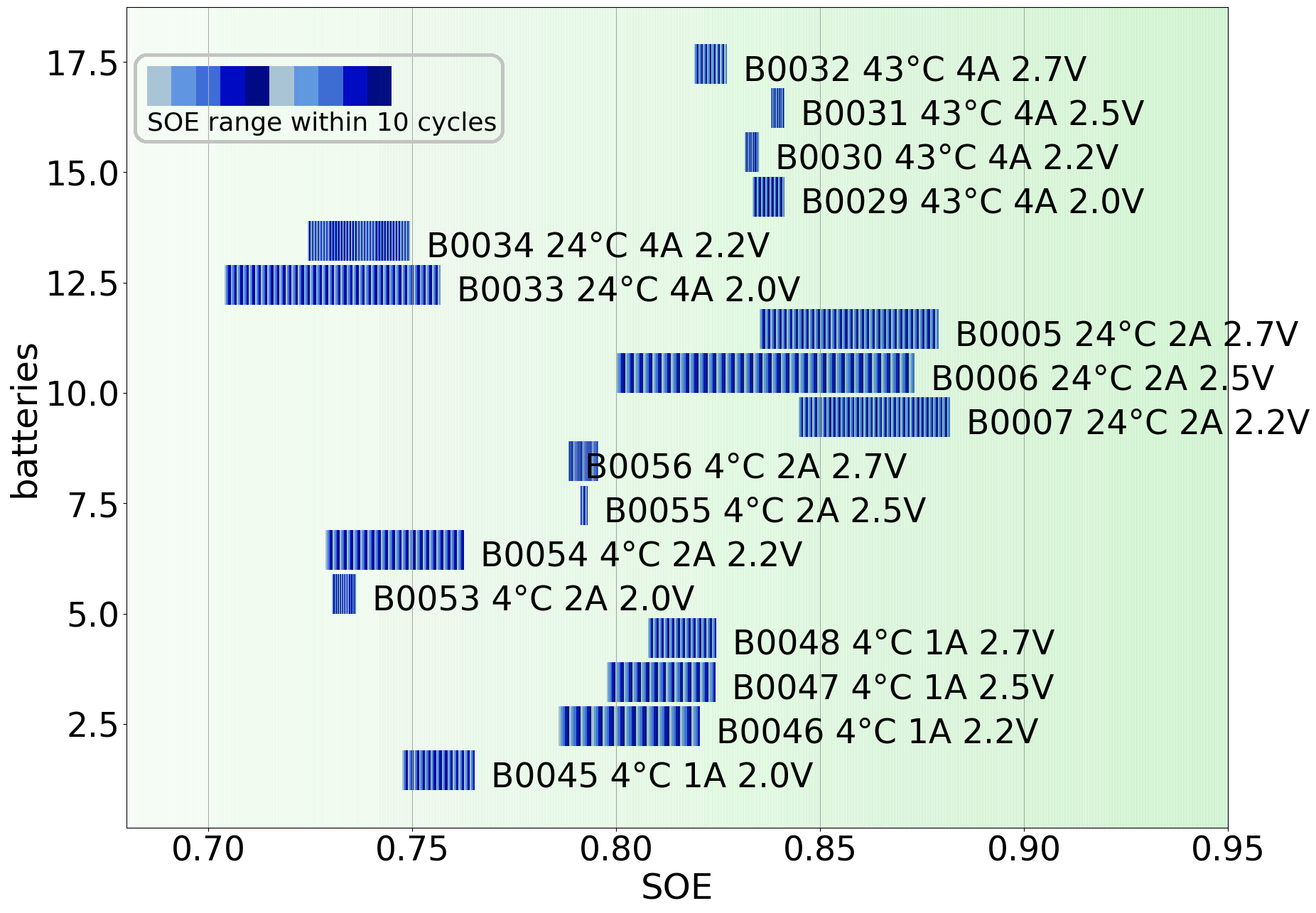}}
		\caption{SOE trends and ranges under various operating conditions.}
        \label{fig:6}
\end{figure}

As the operating conditions of these batteries differ mainly in terms of ambient temperature, discharge current, and cutoff voltage. To collectively visualize the effect of different operating conditions on the energy efficiency of the battery, a summary of the results is shown in \figref{fig:6}. 

As shown in \figref{fig:6} (a), a straight line represents a specific operating condition during the battery life cycle. Longer lines indicate a longer battery life. The blue, green, and red lines correspond to ambient temperatures of 4°C, 24°C, and 43°C respectively. A thinner line indicates a lower discharge current, whereas a thicker line indicates a higher discharge current, and a thick line with edge indicates discharge current of 4A. Dark colors represent deep cutoff voltages, and light colors are for high cutoff voltages. 

\figref{fig:6} (b) shows the SOE ranges of the batteries. Each horizontal bar represents an SOE range from a particular test case. Meanwhile, the rest of the batteries age under constant operation. Right-handed bars have a higher SOE, whereas left-handed bars have a lower SOE. At the beginning of the test, the SOE is located on the right side of the bar. As the battery ages, the SOE shifts to the left. Longer bars indicate a battery with a wider range of SOE during the test, which means these batteries show a more obvious decline in SOE during the test.

\section{Analysis and Discussion}

\subsection{SOE trends and ranges under different operating conditions}

The test schema specifies that EoL conditions occur when battery capacity drops below a certain level, at which point the test is terminated. As a result, the batteries undergo a variation in the number of cycles. Depending on the rate at which battery capacity is degraded, SOE trajectories vary in length. Among all the constant operating conditions included in the data set, batteries has the longest life at 24°C, followed by 4°C, and the shortest at 43°C. However, contrary to SOH, the batteries with the highest SOE were tested at 43°C, 1A. This suggests that discharge current may have a greater effect on SOE than ambient temperature.

Continuous cycling decreases the energy efficiency of most batteries. 
Batteries, which is discharged at 4°C 2A current, despite having a relatively long RUL, 
had a relatively lower SOE than those discharged at 4°C 1A. It is possible that the higher discharge current may have contributed to an extended RUL, but resulted in a suppressed SOE for batteries at extra low temperatures. It is interesting to note that these batteries suffer severely from lower cutoff voltages in terms of SOE at 4°C ambient temperature. The SOE of batteries discharged at 4°C 1A with a voltage of 2.0V and 2.2V has a value of approximately 0.75, while other batteries of the same group with a relatively higher cutoff voltage, have a value of approximately 0.8.

Cycled batteries at 43°C exhibit rapid degradation in terms of SOH, but their SOE appears to be least affected by discharge current and cutoff voltage. 

When tested at 24°C with a 2A discharge current, batteries exhibit a long RUL and a high SOE. In these batteries, the cutoff voltage appears to have a mitigating effect on SOE, and RUL and SOE may be affected by differences in the manufacturing process. When the batteries were tested with 4A at the same temperature, their SOE was again suppressed by the discharge current. It is noteworthy that the data set does not include the 24°C 1A test case. Based on all the results in the data set, it appears that this approach will result in higher efficiency and RUL.

The range of SOE for these batteries is determined by the slope of the trend and the number of cycles. Batteries that have a relatively long RUL and a high tendency to degrade have a longer SOE range. Batteries operating at 24°C 2A have a high initial SOE and a wide SOE range. These characteristics indicate that the batteries' energy efficiency is relatively good at the beginning of the test and decreases as they age. In contrast, batteries that operate at 43°C have a narrow SOE range, primarily due to their short RUL. 4°C 2A batteries have little tendency to decrease with a higher cutoff voltage, so they also have a narrow SOE range.

\subsection{Influence of operating conditions on SOE}

\begin{figure}[htbp]
    \centering  
    \includegraphics[scale=0.3]{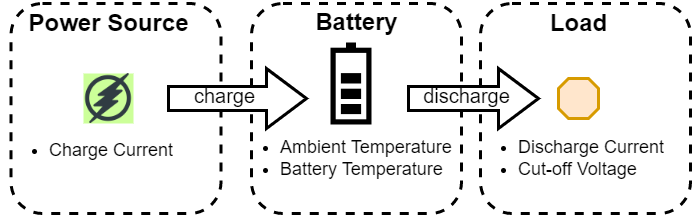}
    \caption{The operating conditions in battery operation scenario.}
    \label{fig:7}
\end{figure}

Operating conditions have a significant impact on the efficiency of a battery. Based on the data set used in this study, ambient temperature, cutoff voltage, charge current, and discharge current were all considered operating conditions, as shown in \figref{fig:7}. Different combinations of ambient temperature, cutoff voltage, and discharge current are tested on different groups of batteries. However, the charge current for all batteries is the same.

\begin{figure}[htbp]
    \centering 
    \includegraphics[scale=0.25]{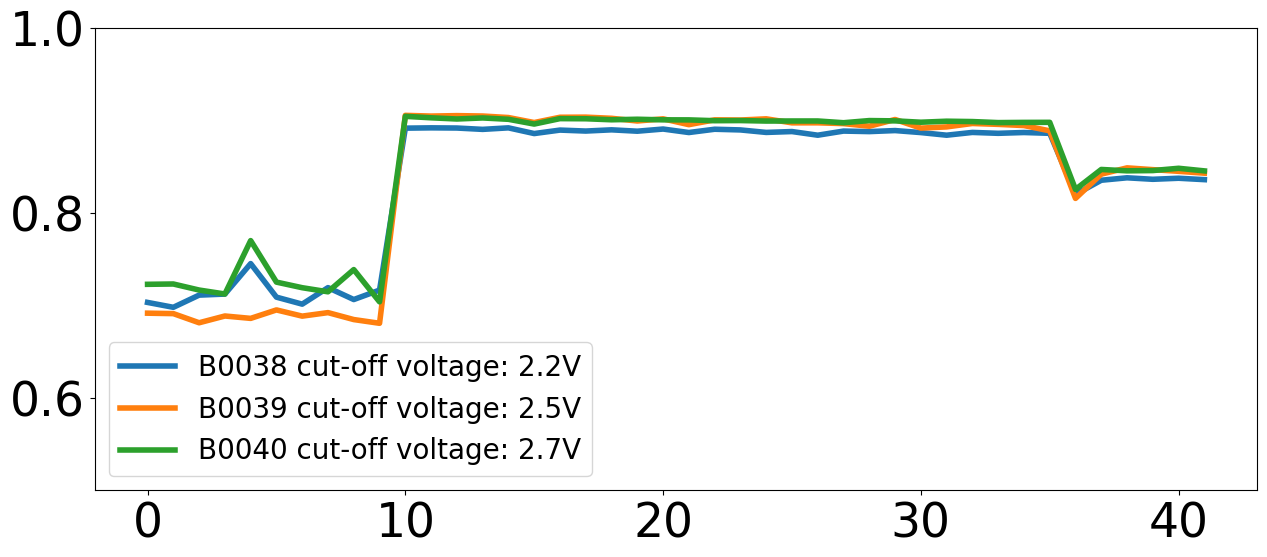}
    \caption{SOE curves for a group of batteries with varying operating conditions.}
    \label{fig:8}
\end{figure}

The B0038, B0039, and B0040 batteries had test conditions that changed during the aging period as opposed to the other groups with constant operating conditions. Initially, the ambient temperature is 24°C, but after the first 11 cycles, it changes to 43°C. Discharge current is initially 4A, is reduced to 1A when the temperature rises to 43°C, and then increases to 2A. Consequently, this battery group consists of three periods of constant operation: 24°C, 4A, 43°C, 1A, and 43°C, 2A. \figref{fig:8} shows the SOE curves after removing the first and last cycles, as well as the 11th cycle (the jump from 24°C to 43°C). There is the lowest efficiency (24°C, 4A) in the first period, and the highest efficiency (43°C, 1A) in the second period. The efficiency decreases during the last period (43°C, 2A), but still outperforms the other group operating at 43°C, 4A. This group of batteries may indicate that energy efficiency can be adjusted under different operating conditions without causing memory effects.

\begin{figure}[htbp]
	\centering  
	\subfigcapskip=-5pt 
	\subfigure[The influence by discharge current.]{
		\includegraphics[width=0.9\linewidth]{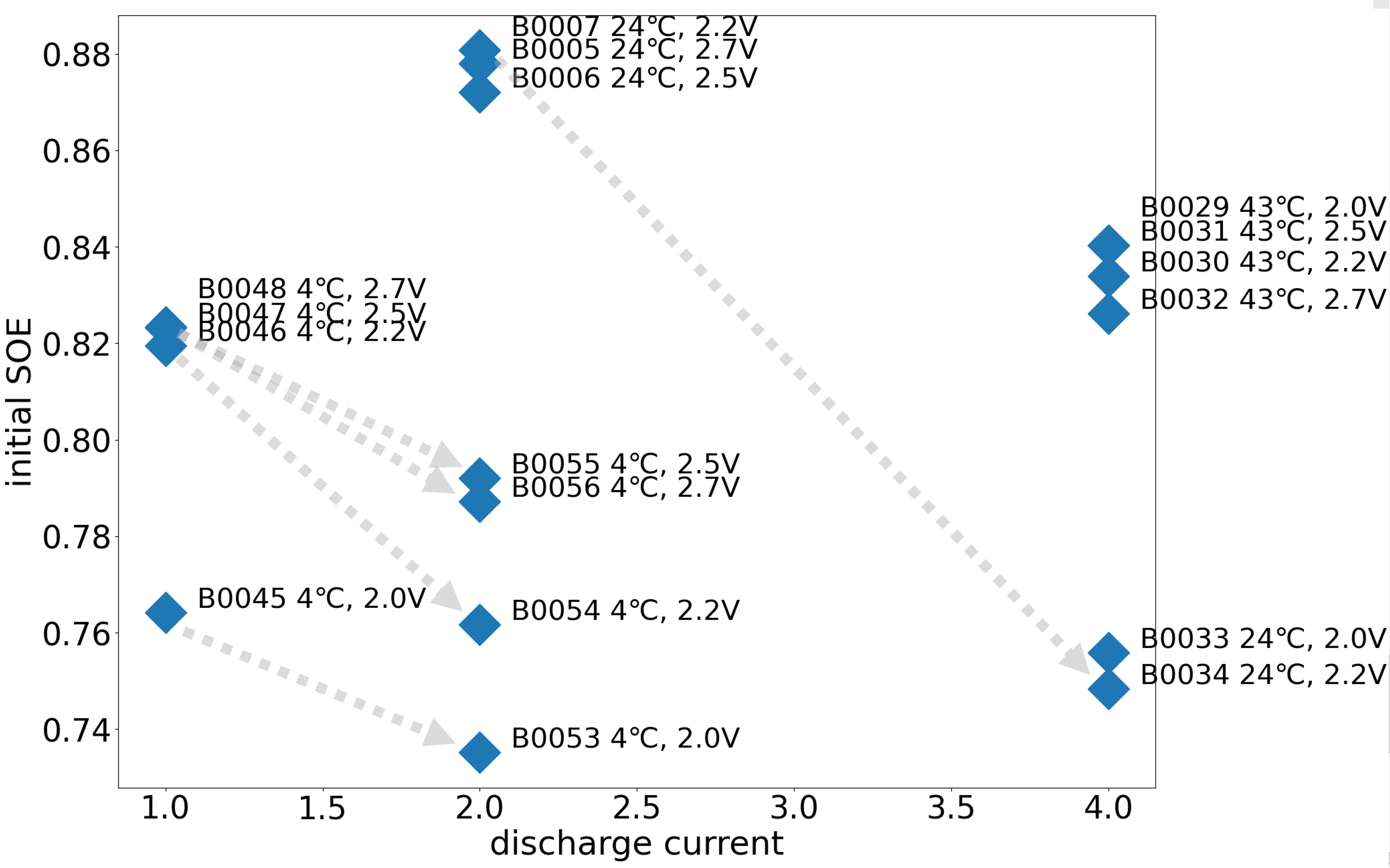}}
	\subfigure[The influence by ambient temperature.]{
		\includegraphics[width=0.9\linewidth]{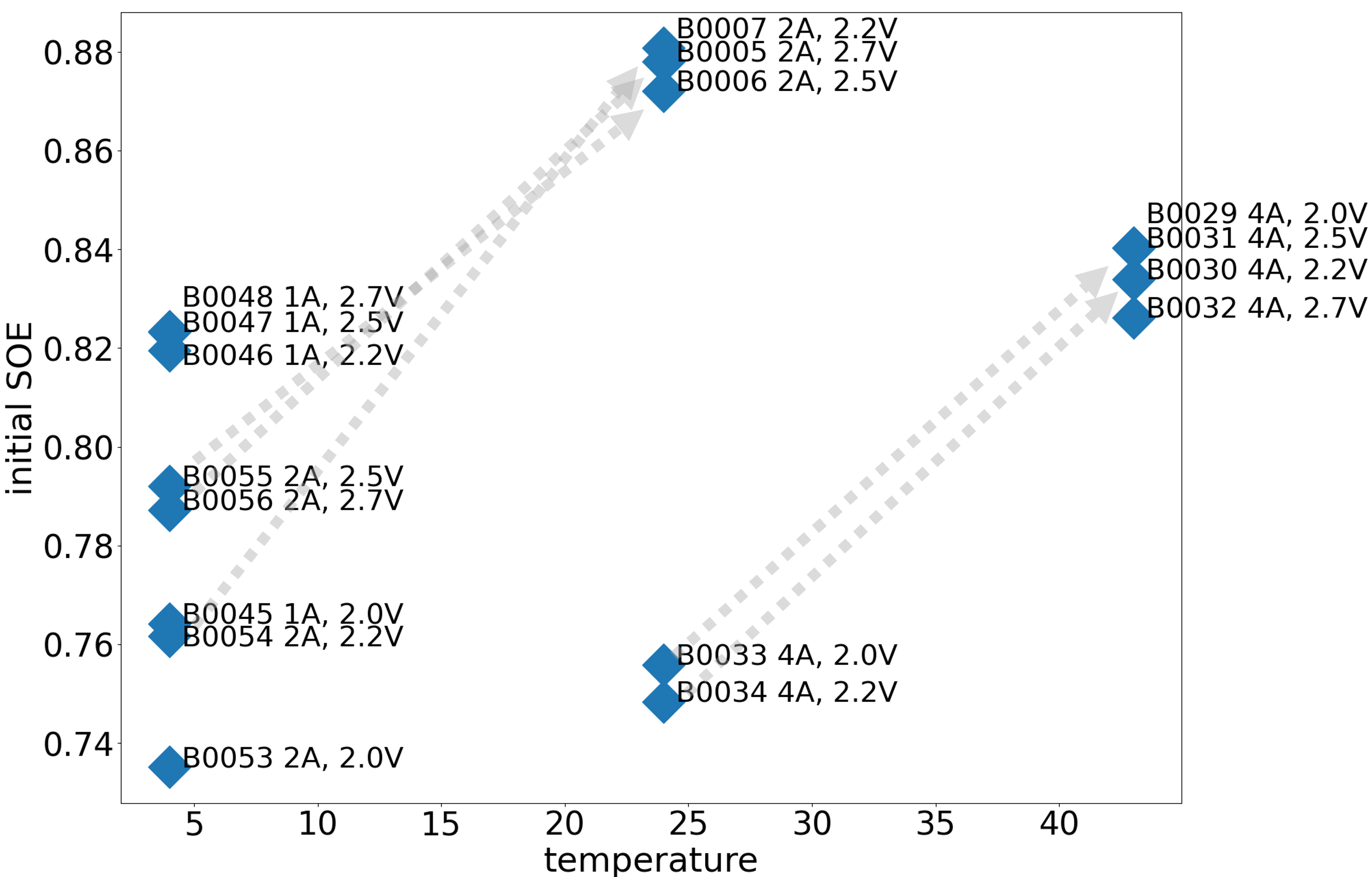}}
	\subfigure[The influence by cutoff voltage.]{
		\includegraphics[width=0.9\linewidth]{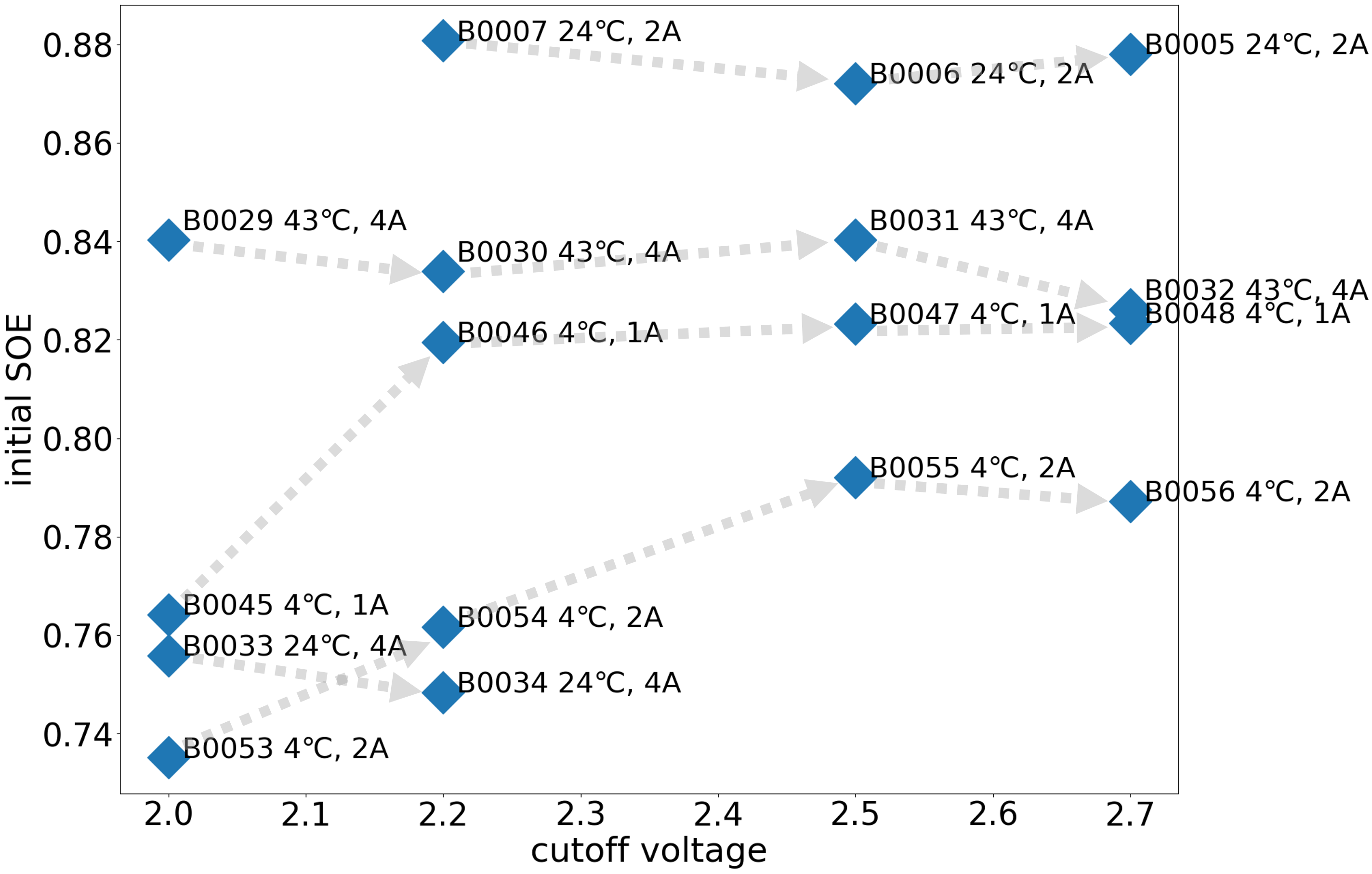}}
	
	\caption{The influence to SOE by operating conditions.}
        \label{fig:9}
\end{figure}

\figref{fig:9} (a) shows that a battery with a lower discharge current is more energy efficient. Higher discharge currents allow a battery to operate at higher power, but they may also negatively affect the battery's SOE. A B0034 discharged at 4A has a SOE of roughly 0.73. On the other hand, the B0007 discharged at 2A has an SOE of more than 0.85, at the same ambient temperature and cutoff voltage. 

Battery performance increases at higher ambient temperatures, as shown in \figref{fig:9} (b) . From 4°C to 24°C, the SOE of a 2A discharged battery improved by almost 0.12; from 24°C to 43°C, the SOE of a 4A discharged battery improved by approximately 0.09.

Discharging batteries at different depths can be achieved by using different cutoff voltages. When a battery is discharged to an extended depth, more energy is released during a single discharge cycle. An increase or decrease in discharge depth, for example, from 2.7V to 2.5V, generally has a limited effect on the SOE, as shown in \figref{fig:9} (c). It should be noted, however, that at ambient temperatures of 4 degrees or 24 degrees, 2.2V or 2.0V cutoff voltages will result in a significantly lower SOE than those at higher cutoff voltages. Using the B0029 as an example, at a depth of discharge of 2.0V, the SOE is only 0.76; other batteries in the group, the B0032 and B0032, with cutoff voltages of 2.5V and 2.7V, have SOEs exceeding 0.86.

\subsection{Implications and suggestions}

Based on NASA's experimental data, we identified some interesting observations and patterns concerning battery energy efficiency. Some of them make sense intuitively, while some are somewhat surprising, nonetheless in many cases they are quite different from the currently existing degradation patterns for SOH or SOC. Clearly, it would be worthwhile to delve further into the reasons underneath these findings, but these observations and patterns themselves can already serve as valuable references when it comes to using these batteries and designing efficient-aware BESSs in the future.

As expected, SOE also gradually degrades just like SOH as the battery ages, but at a much slower pace: when operated under the same conditions, the overall energy efficiency degradation is rarely greater than 4\% throughput its lifespan when its SOH reaches the 70\% stopping point. In comparison, changes in operating conditions on the other hand have a much larger impact on energy efficiency that can easily result in more than 20\% difference in SOE. What is more, SOE does not seem to exhibit a significant `memory effect': when the operating conditions of a battery changes, in the following cycles its SOE will change immediately and reaches a comparable level as if it has been operated under those conditions all along, irrespective of its previous operating conditions.

These behaviors of SOE suggested that, old batteries that are currently considered unusable due to capacity loss, may actually still be useful efficiency-wise, since under the favorable operating conditions, their energy efficiency will not be much different from that of new batteries. There is still considerable potential for these batteries to provide efficient energy caching for renewable energy, and capacity loss can always be compensated by quantity. In this way, retired batteries can largely be reused for storage purposes such as caching the renewable energy in a cost-effective, more environmentally friendly and efficient manner. 
\par
Comparing to aging, operating conditions are much larger factors that determine the energy efficiency of a battery. For BESS, the performance of batteries varies due to production deviations, inhomogeneous aging, and temperature differences within the cluster. With a battery pack containing batteries with differing SOEs and SOHs, the overall energy efficiency of BESS may be adversely affected. In order to maximize the performance and efficiency of a battery pack, active balancing must be implemented in order to equalize the performance of these batteries.

\dgli{First, intensive discharging at high current always has a negative impact on energy efficiency, therefore to improve the overall efficiency of a BESS, we should carefully control the discharge current of individual batteries and keep them balanced and at a lower level.} 
On the other hand, although it has been well received that both low and high temperatures have a detrimental effect on battery capacity or SOH, the effect of higher ambient temperatures on SOE, however, is essentially positive. In other words, the battery's energy efficiency can be improved when the temperature is kept warm within its normal operating range. Low temperatures have a further negative effect on SOE: deep discharge of a battery under low temperature conditions results in a rapid drop in energy efficiency; while as at higher temperatures, SOE is generally not sensitive to deep discharge towards low cut-off voltages.

\dgli{The fact that SOE does not have a significant memory effect can also inspire us when designing battery management algorithms for BESS. For example, we can be more tolerant to short-term in-efficient operations, such as large-current and deep discharging in cold environment, for emergency situations without worries of long-term damage on SOE, as long as we can resume favorable conditions after the situation is resolved. This gives us more flexibility to handle demand surges or deal with circumstances in the grid or the solar or wind farms that undergo unusual situations. Of course, because our data is much limited, whether extreme intensive use of the batteries under a lengthy period of unfavorable conditions will do irreversible harm to the energy efficiency still needs further investigations with additional experiments and data. This is also part of our future work.}

\section{Conclusions}

Efficiency of batteries, particularly those used in ESSs, will have a significant impact on power systems. In this study, we proposed SOE as an indicator of the battery's performance, and evaluated the SOE of NCA lithium-ion batteries in the well-known data set.

Our study examined the SOE trends of these batteries under a variety of operating conditions. On the basis of the test of linear trend, we developed a SOE trend model that is applicable to this data set. We also discussed how different operating conditions under different scenarios affected the SOE of the batteries. Results of the regression of the SOE trend model have shown that increasing ambient temperature and decreasing discharge current have a positive impact on energy efficiency. The results also indicated that at low ambient temperatures, the battery's energy efficiency may be significantly reduced when operating at an extremely low cutoff voltage. As both aging and operating conditions have an impact on energy efficiency, BMS controllers should monitor the parameters of each battery, including terminal voltage, ambient temperature, charging and discharge current, so as to ensure performance for energy efficiency. 

There are a number of limitations to this study that should be acknowledged. The size of the data set currently used restricts further analysis of the impact of a wide range of operating conditions on SOE trends. As our analysis is based on NCA lithium-ion batteries, it may be necessary to develop more complex models to estimate the energy efficiency of different lithium-ion batteries under a broader range of scenarios.

    










\printcredits

\bibliographystyle{elsarticle-num}

\bibliography{cas-refs}

\bio{}
\endbio

\endbio

\end{document}